\documentclass[prl,aps,twocolumn,nofootinbib,floatfix]{revtex4}
\usepackage{graphicx}

\def\be{\begin{equation}}
\def\ee{\end{equation}}
\def\ba{\begin{eqnarray}}
\def\ea{\end{eqnarray}}

\def\tr{{tr}}
\def\lng{{l}}
\def\D{\Delta}
\def\x{{\bf x}}
\def\k{{\bf k}}
\def\Re{\mbox{Re\,}}
\def\Im{\mbox{Im\,}}
\def\cA{{\cal A}}

\def\ome{\omega}

\begin{document}
\title{A universal holographic prediction for quantum-critical dynamics}
\author{Sergei Khlebnikov}
\affiliation{Department of Physics and Astronomy, Purdue University,
West Lafayette, IN 47907, USA}
\begin{abstract}
We consider decay of an initial density or current perturbation at finite 
temperature $T$ near a quantum critical point with emergent Lorentz invariance.
We argue that decay of perturbations with wavenumbers $k \gg T$ (in natural 
units) is a good testing ground for
holography---existence of a dual gravitational description---in experimentally 
accessible systems. The reason is that, computed holographically, the decay rate 
at large $k$ depends only on the
leading correction to the metric near the boundary, and that is quite universal.
In the limit of zero detuning (when the temperature is the only dimensionful 
parameter), the result is a scaling law for the decay rate, with the exponent that
depends only on the dimensionality. We show that this follows from an analytical 
argument and is borne out by a numerical study of quasinormal modes.
\end{abstract}
\maketitle

Recently, much attention has been attracted to real-time dynamics in conformal field
theories (CFTs), especially those capable of describing interacting
quantum critical points of interest in condensed matter physics
\cite{Damle&Sachdev}. When the interactions are strong, 
the quasiparticle picture is not applicable
for interpretation of the zero-temperature ($T = 0$) spectrum,
and one needs new ideas on how to think
about relaxation and transport in the $T \neq 0$ fluid. 
One such idea, originally developed in the context of
string theory \cite{Maldacena:1997,Gubser&al,Witten:1998} (for a review, see Ref.
\cite{Aharony&al}), is holography, according to which certain types
of strongly-interacting CFTs have an equivalent (dual) description in terms of string
theory or gravity 
in an extra-dimensional space. This 
has been applied to a study of quantum critical 
transport in a CFT with a known gravity dual in Ref.~\cite{Herzog&al}. 
For a review of holography from the condensed matter perspective, see
Ref.~\cite{Hartnoll&al}.

Most of the current dual gravity proposals are for gauge theories; in addition,
for gravity to be semiclassical, the number $N$ of colors in the gauge theory must be large
\cite{Maldacena:1997}.
One may ask if holography-based methods can at all 
be extended to a broader class
of theories. An existing approach is
to think of holography as a method of analytical 
continuation of results obtained in the Euclidean (imaginary time) signature to real 
time \cite{Katz&al}.
In that approach, one fits the Euclidean data, assumed available through other means, 
by a gravity dual with a few free parameters and then uses the result for 
making predictions for quantities measurable in real time. For example, it has
been found \cite{Katz&al} that a good fit to the Euclidean conductivity at the 
O(2) fixed point in (2+1) dimensions can be 
obtained with only one additional field, corresponding to a relevant operator
in the CFT. A holographic model in which this field becomes dynamical has been constructed
in Ref.~\cite{Myers&al}.

A different approach, which we adopt here, is to forgo reconstruction of the full
extra-dimensional  (``bulk'') geometry and focus instead on ``seeing''
a region near the boundary. Consider, as we do throughout this paper, a CFT that has 
a (possibly emergent) Lorentz invariance. Then, according to the holographic 
correspondence \cite{Maldacena:1997,Gubser&al,Witten:1998}, the near-boundary geometry 
is asymptotically
anti-de Sitter (AdS). At temperature 
$T \neq 0$, there is a small metric correction, due to a black hole 
residing in the AdS bulk. 
The question we ask is whether it is possible to detect the metric correction by making
measurements (ultimately, in the laboratory) that are defined on the CFT side.
The answer is non-obvious because, as just discussed, there are 
additional bulk fields that may have nontrivial profiles. It is true, for instance, that
a scalar corresponding to a relevant operator (one with a scaling dimension 
$\D < d$ in a $d$-dimensional CFT) vanishes at the boundary, but so does the metric
correction. It is not
obvious if one can detect the latter without worrying about the former.

The measurements we have in mind are of the rate of 
relaxation of an initial perturbation. We consider CFTs (such as the already mentioned 
O(2) fixed point) that have a conserved current density, $J^\mu$, and study relaxation of
an initial nonzero average $\langle J^\mu \rangle$ 
to the equilibrium (zero) value. The index $\mu$ runs over all $d$ values, 
$\mu = 0,\dots d-1$; in particular, the temporal component $J^t \equiv J^0$ is
the corresponding charge density. In linear response
theory, the process is governed by singularities of the retarded Green function of 
$J^\mu$ in the complex frequency plane. The setup then is mostly the same as in
Refs.~\cite{Herzog&al,Miranda&al,dispersing,Katz&al}, 
except that we focus on a particular kinematic region: we consider perturbations whose
wavenumbers $k$ are large, $v k \gg T$, where 
$v$ is the effective ``speed of light'' characterizing a Lorentz-invariant
CFT. The corresponding complex frequency 
is near the light cone, $\ome \approx v k$. 
In what follows, we use units in which $v = 1$. 

For most of the paper, we consider the case when $T \neq 0$ is the only deformation
from exact criticality,
meaning that the coupling constants are all at their critical values---the
so-called zero detuning limit.
More precisely, we consider cases in which the metric
near the boundary can be brought to the from
\be
ds^2 = \frac{R^2}{\lambda_T^2 u^2} \left[ -f(u) dt^2 + d\x^2  \right] 
+ \frac{R^2 \ell(u)}{u^2} du^2 \, ,
\label{ds2}
\ee
where $u$ is the radial coordinate such that the boundary is at $u = 0$, 
$d\x^2$ is the flat Euclidean metric in $(d-1)$ dimensions, $R$ is the
AdS curvature radius, and $f(u)$ has the following small-$u$ asymptotics:
\be
f(u) = 1 - u^d + \dots \, .
\label{fu}
\ee
Of $\ell(u)$, we require only that $\ell(u) \to 1$ at $u \to 0$. The parameter $\lambda_T$
is a result of rescaling $u$ to make the $u^d$ correction in (\ref{fu}) have the unit
coefficient. For a black hole with a planar horizon (a black brane) in otherwise empty
AdS, $f(u) = 1/\ell(u) = 1 - u^d$ for all $u < 1$. We refer to this space as (planar)
AdS-Schwarzschild (AdSS). In that case, $\lambda_T$ is inversely proportional 
to the Hawking temperature $T_H$ of the black hole, so even in the more general case
(\ref{fu}) we refer to it as the thermal wavelength.

The key is that, although we allow additional fields to deform the metric in the bulk,
in Eq.~(\ref{fu})
we assume that near the boundary the metric retains the AdSS form, up to and including the
$u^d$ correction for $f(u)$. The length $\lambda_T$ does not have to be related to 
$T_H$ in exactly the same way as for AdSS, but zero detuning implies that it remains
inversely proportional to it.  

To see why Eq.~(\ref{fu}) is plausible, consider for 
instance the bulk stress tensor of a scalar corresponding to an operator of dimension
$\D$. By the standard AdS/CFT dictionary \cite{Gubser&al,Witten:1998}, 
the asymptotics of the scalar 
near the boundary for zero detuning is $\phi(u) \sim u^\D$, so the leading term in
the stress tensor there is
\ba
\Theta_{tt} & = &  \lambda_T^{-2} (\D^2 + M^2) \cA u^{2\D -2} \, , \label{tt} \\
\Theta_{uu} & = & (\D^2 - M^2) \cA  u^{2\D - 2} \, , \label{uu}
\ea
where $M$ is the mass of the scalar, in units of $1/R$, and $\cA > 0$ is a constant. 
We can use (\ref{tt}) and (\ref{uu}) to find corrections to the
functions $f$ and $\ell$ in (\ref{ds2}). Because of the relation $\D (\D - d) = M^2$
\cite{Gubser&al,Witten:1998}, the expression (\ref{tt}) for $\Theta_{tt}$ vanishes at
$\D = d/2$. Otherwise, solution of the Einstein equations near the boundary shows that,
while $\ell(u)$ receives a correction proportional to $u^{2\D}$, the corresponding term
in $f(u)$ cancels out. For $\D = d/2$, the stress tensor (\ref{tt})--(\ref{uu})
is formally ``resonant'' with
the $u^d$ term in the metric functions, so one might expect a logarithmic correction.
But, because in this case $\Theta_{tt} = 0$, there is in fact none.
Thus, the form (\ref{fu}) is quite generic.

In what follows, we argue that, as a consequence of (\ref{fu}), the relaxation rate
of $\langle J^\mu \rangle$ at $k \gg T$, in both longitudinal and transverse 
channels, is governed at large times by a universal scaling law that depends only
on the dimensionality.\footnote{The scaling exponent in it is the same as that found 
in Refs.~\cite{Festuccia&Liu,Morgan&al,Fuini&al} for large-$k$ 
scalar and gravitational perturbations in planar AdSS. I am grateful to S. Hartnoll 
for bringing Ref.~\cite{Festuccia&Liu} to my attention. \label{foot}} 
For example, for a Lorentz-invariant CFT
in $d =3$ dimensions, the rate in either channel is of the form
\be
- \Im \ome \approx C T^{6/5} k^{-1/5} \, ,
\label{rate}
\ee
where $C$ is a numerical constant. The constants, $C = C_{\tr}$ or $C_{\lng}$,
are different for the two channels, and their ratio (although not each constant
by itself) is universal. A measurement of
the relaxation rate at large $k$ thus provides a test
(a necessary condition) of whether a given critical point has a holographic dual.
Towards the end of the paper, we discuss prospects
for a such a measurement. 

We start with an analytical
argument in favor of Eq.~(\ref{rate}) and then confirm the result by a numerical 
calculation. 

The thermodynamic variable conjugate to $J^\mu$ is a vector field $A_\mu$.
By the standard dictionary \cite{Gubser&al,Witten:1998}, in holography it becomes
the boundary value of a Maxwell gauge field, $A_M$, propagating in the 
$(d+1)$-dimensional bulk. The Maxwell equation, in the presence of 
additional (charge neutral) fields, is 
\be
\partial_M \left[ h(u) \sqrt{-g(u)} g^{MN}(u) g^{PQ}(u) F_{NQ}(t, \x, u) \right] = 0 \, ,
\label{eqm} 
\ee
where all indices run over $d+1$ values, $g_{MN}$ is the metric tensor,
$F_{NQ} = \partial_N A_Q - \partial_Q A_N$, and
$h$ describes the coupling to the additional fields. For example, in the 
scenario of Ref.~\cite{Katz&al}, there is one additional field, a scalar $\phi(u)$,
and $h(u) = 1 + \alpha \phi(u)$. At zero detuning, $\phi$ vanishes
at $u \to 0$ as $u^\Delta$ with $\Delta \approx 3/2$. Here, we require only that 
$h(u) \to 1$ when $u \to 0$. 

Because the metric (\ref{ds2})
does not depend on $t$ and $\x$, we can look for solutions to (\ref{eqm}) 
in the form $e^{-i\ome t + i {\bf k} \cdot \x}$ times a radial dependence. 
The AdS radius 
$R$ scales out of the Maxwell equation. In addition, if we measure 
$\ome$ and $\k$ in units of $1/ \lambda_T$, so does $\lambda_T$. So, in what follows,
$g_{MN}$ will denote the metric tensor corresponding to (\ref{ds2}) with $R=\lambda_T = 1$.

Next, we impose the gauge condition $A_u = 0$ and obtain separate equations for the 
longitudinal and transverse components of $A_\mu$. 
The derivation is standard
(see, for example, Ref.~\cite{Herzog&al}).
Transverse solutions have
$A_t = 0$ and $\sum_i \partial_i A_i = 0$ ($i = 1, \dots, d - 1$ labels the spatial
directions in the boundary theory).  Their radial dependencies,
which we call simply $A(u)$, satisfy
\be
\partial_u \left[ w_\tr(u) g^{uu}(u) \partial_u A \right] + w_\tr(u) g^{xx}(u) 
\left[ \frac{\ome^2}{f(u)} - k^2  \right] A = 0 \, ,
\label{tr}
\ee
where $w_\tr \equiv h \sqrt{-g} g^{xx}$, and $g^{xx} = u^2$. For the longitudinal
modes, it is convenient to work with the ``electric'' field 
$Z \equiv w_\lng g^{uu} \partial_u A_t$, where $w_\lng \equiv -h \sqrt{-g} g^{tt}$. 
The mode equation for $Z$ is
\be
\partial_u \left[  \frac{1}{w_\lng(u) g^{xx}(u)} \partial_u Z \right]
+ \frac{1}{w_\lng(u) g^{uu}(u)} \left[ \frac{\ome^2}{f(u)} - k^2  \right] Z = 0 \, .
\label{long}
\ee
The values of $A_\mu$ at $u = 0$ deform the system away from criticality and at zero
detuning must vanish. This produces the following boundary conditions 
for the transverse $A$ and the longitudinal $Z$ at $u\to 0$:
\ba
A(u = 0) & = & 0 \, , \label{bcA} \\
\lim_{u \to 0} u^{d-3} \partial_u Z(u) & = & 0 
\label{bcZ} \, .
\ea
The latter is a consequence of the Gauss law and the asymptotics
$w_{\lng} g^{xx} \to u^{3-d}$.

Poles of the retarded Green function of $J_\mu$ coincide with the quasinormal
modes of the field $A_M$, defined as solutions to
the mode equations subject to zero detuning conditions at $u \to 0$ and the infalling 
boundary condition at the horizon \cite{Horowitz&Hubeny,Nunez&Starinets}. 
How should we think about the infalling condition
when we only have access to the 
near-boundary geometry? A useful analogy is with the 
escape of a photon from a braneworld into a bulk black hole \cite{BBH}.
As in that case, one can define the escape radius $u_e$ by the condition
that at $u \sim u_e$ solutions, say, to Eq.~(\ref{tr}) become oscillatory 
(so one can identify waves that are outgoing and incoming with respect to 
the boundary).
If we think of (\ref{tr}) as the Schr\"{o}dinger equation for an equivalent particle,
the wave that is outgoing at $u > u_e$ describes a particle that propagates
away from the boundary. Assuming there are no ``bumps'' 
at larger $u$ that would reflect
the particle back, we can impose the outgoing boundary condition at $u$ somewhat larger
than $u_e$, instead of the horizon. If $u_e \ll 1$, the resulting boundary problem
will be entirely within the domain of applicability of the small-$u$ limit (\ref{fu}).

We now describe an
analytical argument in favor of the scaling law (\ref{rate}) (and
similar ones in other dimensionalities).  
Consider the mode equation (\ref{tr})
for $\ome \approx k$ and large $k$. Assuming $u_e \ll 1$, to be checked 
{\it a posteriori}, take the small $u$ limit. In the potential (second) term,
the metric correction gets multiplied by the large $k^2$: we have
\be
\frac{\ome^2}{f(u)} - k^2 \approx k^2 u^d + m^2 \, ,
\label{pot}
\ee
where $m^2 \equiv \omega^2 - k^2$. The enhancement by $k^2$ in (\ref{pot}) is the
reason why we will be able to see the $u^d$ correction to the metric even if
the rest of the geometry is not known precisely.
There is no such enhancement in any other factors,
so we can set the remaining metric components to their asymptotic limits and
$h(u) \to 1$. A change of variables from $u$ to $\rho = 2 k u^{d/2 + 1} / (d + 2)$ and
from $A$ to $B$, where $A(\rho) = \rho^\nu B(\rho)$ and $\nu = (d -2) / (d+2)$, brings 
the resulting (approximate) equation to the form
\be
B'' + \frac{1}{\rho} B' - \frac{\nu^2}{\rho^2} B + B = - \frac{m^2}{k^2 u^d} B \, ,
\label{bessel}
\ee
which is the Bessel equation with an additional potential \cite{BBH}. Purely outgoing waves
(resonances) are poles of the corresponding scattering amplitude analytically
continued to complex $m^2$ (at real $k$). At fixed $\rho$, $u$ scales with 
$k$ as $k^{-2/(d+2)}$. Eq.~(\ref{bessel}) then shows that the locations of the poles
can depend only on the combination $m^2 k^{-4/(d+2)}$, so $m^2 \sim k^{4/(d+2)}$ or,
equivalently,
\be
- \Im \ome \sim  k^{(2-d)/ (2 + d)}  \, .
\label{Im_ome}
\ee
For this value of $m^2$, solutions to (\ref{bessel}) start to oscillate at 
$\rho \sim 1$, so the condition $u_e \ll 1$ for the
escape distance is satisfied.
The parallel estimate for the longitudinal channel differs in details but gives
the same result (\ref{Im_ome}). In both channels, then,  
the scaling exponent is the same as that found 
in Refs.~\cite{Festuccia&Liu,Morgan&al,Fuini&al} 
for large-$k$ scalar and gravitational modes in the full planar AdSS.

We now turn to numerical results. We test universality of the near-boundary limit by
using the full planar AdSS metric, i.e., Eq.~(\ref{ds2}) with $f(u) = 1/\ell(u) = 1 - u^d$ 
for all $u < 1$, and different forms of $h(u)$. If the results do not
depend significantly on the choice of $h(u)$, we can infer that we have in fact used only 
the small-$u$ limit of the theory. 
For the full AdSS, we can impose the boundary condition directly at the horizon,
at $u\to 1$, where
(using the transverse channel as an example)
\be
A(u) = (1 - u)^s \left[ 1 + c_1 (1 -u) + c_2 (1 - u)^2 + \dots \right] ,
\label{bc_hor2}
\ee
$s = -i\ome / d$, and the constants $c_1$, $c_2$, $\dots$ 
are determined from Eq.~(\ref{tr}). We employ the shooting method, starting at 
$u = 1 - \delta$ with a small $\delta$ and computing $A$ and $\partial_u A$ at
that $u$ from Eq.~(\ref{bc_hor2}). A similar method was used
for computation of quasinormal frequencies (in a different context) in Ref.~\cite{shock}.
The number of terms in the bracket in (\ref{bc_hor2}) should
be sufficient to reject the second, $(1- u)^{-s}$, solution 
(assuming $2s$ is not an integer).
It depends on how large the real part of $s$ is \cite{shock}. For example,
for $\Re s > -1/2$, keeping only the linear term is sufficient; for
$- 1 < \Re s \leq -1/2$, one needs also the quadratic term, etc. 

\begin{figure}[t]
\begin{flushright}
\includegraphics[width=3.25in]{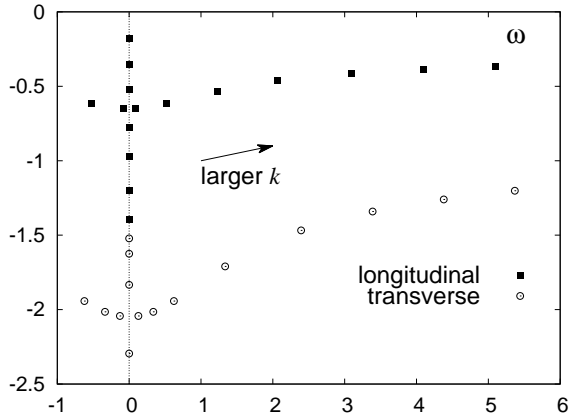}
\end{flushright}
\vspace{0.2in}                                          
\caption{The quasinormal frequencies with the smallest $|\Im{\ome}|$ 
(one for each channel) as they travel in the complex $\ome$ plane with changing
(real) $k$. These results are for $d = 3$, $h(u) \equiv 1$.
}
\label{fig:rlocus}  
\end{figure}

To orient ourselves in the complex $\omega$ plane, we first look at the locus of
quasinormal frequencies (poles of the corresponding response functions) for
small to moderate values of $k$. Fig.~\ref{fig:rlocus} shows the frequencies 
closest to the real axis for $d = 3$ and $h(u) \equiv 1$ (i.e.,
no coupling to additional fields). At moderate $k$, there are pairs of these
located symmetrically about the imaginary axis. At smaller $k$, they collide on the
imaginary axis and remain there as $k$ decreases further \cite{Miranda&al,dispersing}.
In the longitudinal channel, the upper one then
proceeds towards the origin, where it becomes the diffusive pole familiar 
from Refs.~\cite{Nunez&Starinets,Herzog&al,BBH}. 

For $d=3$ and $h\equiv 1$, the Maxwell field in planar AdSS enjoys electric-magnetic 
duality, as a consequence of which poles of the longitudinal response (leaving aside the
one at $\ome = 0$ for $k = 0$) coincide with zeros
of the transverse one, and vice versa \cite{Herzog&al}. In particular, at $k=0$,
when the two response functions coincide, 
neither can have any poles (or zeroes).
We see that this occurs by the lower longitudinal pole and the upper transverse one
moving toward each other and annihilating at some value of $\ome$.
Numerically, the latter is very close to $\ome = -1.5 i$, which is consistent with the
results of Refs.~\cite{Miranda&al,dispersing}.

\begin{figure}[t]
\begin{center}
\includegraphics[width=3.25in]{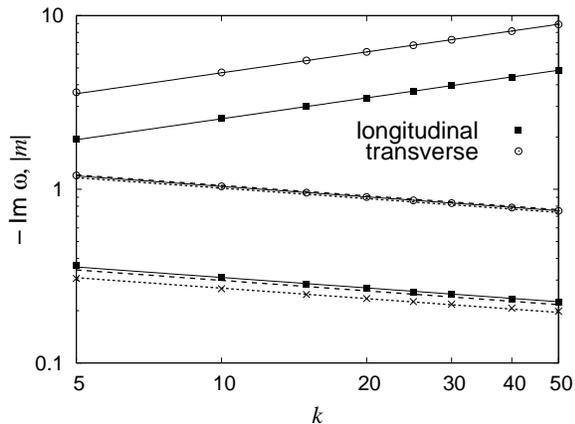}
\end{center}
\vspace{0.2in}                               
\caption{Scaling of $-\Im\ome$  and $|m|$ with $k$, at large $k$, 
for the poles closest to the real axis in $d = 3$.
The squares and circles are data points from the numerical calculation for
$h(u) \equiv 1$, and the solid lines are $k^{-1/5}$ fits to $-\Im\ome$ and
$k^{2/5}$ fits to $|m|$.
The dashed lines are $k^{-1/5}$ fits to $-\Im\ome$ for
$h(u) = 1 + \alpha u^\D$, with long dashes for $\D = 1.5$, $\alpha = 0.6$, 
and short dashes for $\D = 0.5$, $\alpha =1$. The crosses are data points for the 
longitudinal channel in the latter case.
}
\label{fig:scaling}  
\end{figure}

In the present work, our main interest is the behavior of the poles for
large $k$, when they move beyond the right edge of Fig.~\ref{fig:rlocus}. 
In Fig.~\ref{fig:scaling}, we show the corresponding values of $\Im\ome$, together
with $k^{-1/5}$ fits suggested by Eq.~(\ref{Im_ome}). We see that the fits work well.
So do the $k^{2/5}$ fits for $|m|$.
In the same figure, we compare results for 
$h(u) \equiv 1$ and $h(u) = 1 + \alpha u^\D$ with two choices of $\D$ and $\alpha$:
$\D = 1.5$, $\alpha = 0.6$, as
obtained in Ref.~\cite{Katz&al}
by fitting the Euclidean data for the O(2) fixed
point, and $\D = 0.5$, $\alpha = 1$ ($\D \geq 0.5$ being the unitarity bound 
for a scalar in $d=3$ \cite{Aharony&al}). 

At zero detuning, we can restore the power of $T$ in Eq.~(\ref{rate}) on dimensional
grounds. Because we do not insist on a particular numerical factor in the relation 
between $T$ and
the thermal wavelength $\lambda_T$ that appears in the near-boundary metric,
the constant $C$ in (\ref{rate}) is not universal. The ratio between the constants
for the longitudinal and transverse channels, however, is. From the data for 
$h \equiv 1$ (which we expect to reflect the large $k$ behavior better than those
for a general $h$), we find
\be
C_{\lng} / C_{\tr} = 0.30 \, .
\label{ratio}
\ee
These results are for $d = 3$.
For $d= 4$, Eq.~(\ref{Im_ome}) predicts $- \Im \ome \sim k^{-1/3}$, which is also borne out
numerically.

A candidate system where one may be able to experimentally test the scaling (\ref{rate}) 
is a gas of repulsive bosons in a two-dimensional optical lattice. The Bose-Hubbard model,
commonly used for theoretical description of such a gas,
is expected to undergo a quantum phase
transition from a superfluid to a Mott insulator, with a critical point 
(``tip of the lobe'') in the $d=3$ O(2) universality class \cite{Fisher&al}. 
On a square lattice, a typical ``large'' $k$ is of order $\pi /a$, where $a$ is the
lattice spacing. The sound velocity $v$ near the critical point (for unit filling) is 
$v = 5 J a$, where $J$ is the hopping amplitude \cite{MCStudy}. 
The low-temperature condition, 
under which Eq.~(\ref{rate}) is applicable, becomes $T \ll v k \sim 5 \pi J$. 
For a realistic $J$ of a few nK, this is satisfied for 
$T \sim 1$ nK, which looks feasible.

Obtaining large homogeneous critical regions in experiment is a challenge.
Consider in this context a particular detuning effect---deviation $\delta \mu$ 
of the chemical
potential from the critical value. In holography, this is represented by
a nonzero temporal component, $A_t(u =0)$, of the Maxwell field at the boundary or,
in terms of the bulk electric field $Z$, to a nonzero second (subleading) 
term in the small-$u$ asymptotics $Z(u) = a_1 + a_2 u + \dots$. Since neither term in
the asymptotics affects the correction term in Eq.~(\ref{fu}), the scaling law 
(\ref{rate}) still applies, except that $C$ is now a function of $\delta \mu / T$. 
The ratio $C_{\lng}/ C_{\tr}$ remains at the universal value (\ref{ratio}). Thus, 
spatial variations of the trap potential in experiment may need to be kept small,
but perhaps not prohibitively so.

The author thanks Chen-Lung Hung and 
Martin Kruczenski for discussions. This work was supported in part by the U.S.
Department of Energy (grant DE-SC0019202) and by the W. M. Keck Foundation.


\begin{thebibliography}{99}
\bibitem{Damle&Sachdev}
K. Damle and S. Sachdev,
Phys. Rev. B {\bf 56}, 8714 (1997) [cond-mat/9705206].
\bibitem{Maldacena:1997}
J. M. Maldacena, 
Adv. Theor. Math. Phys. {\bf 2}, 231 (1998) [hep-th/9711200].
\bibitem{Gubser&al}
S. S. Gubser, I. R. Klebanov, and A. M. Polyakov, 
Phys. Lett. B {\bf 428}, 105 (1998) [hep-th/9802109].
\bibitem{Witten:1998} E. Witten, 
Adv. Theor. Math. Phys. {\bf 2}, 253 (1998) [hep-th/9802150].
\bibitem{Aharony&al} O. Aharony, S. S. Gubser, J. Maldacena, H. Ooguri, and Y. Oz, 
Phys. Rept. {\bf 323}, 183 (2000) [hep-th/9905111].
\bibitem{Herzog&al} C. P. Herzog, P. Kovtun, S. Sachdev, and D. T. Son,
Phys. Rev. D {\bf 75}, 085020 (2007) [hep-th/0701036].
\bibitem{Hartnoll&al} S. A. Hartnoll, A. Lucas, and S. Sachdev,
arXiv:1612.07324.
\bibitem{Katz&al} 
E. Katz, S. Sachdev, E. S. S{\o}rensen, and W. Witczak-Krempa,
Phys. Rev. B {\bf 90}, 245109 (2014) [arXiv:1409.3841].
\bibitem{Myers&al}
R. C. Myers, T. Sierens, and W. Witczak-Krempa,
JHEP {\bf 1605}, 073 (2016) [Addendum: JHEP {\bf 1609}, 066 (2016)]
[arXiv:1602.05599].
\bibitem{Miranda&al}
A. S. Miranda, J. Morgan, and V. T. Zanchin,
JHEP {\bf 0811}, 030 (2008) [arXiv:0809.0297].
\bibitem{dispersing}
W. Witczak-Krempa and S. Sachdev,
Phys. Rev. B {\bf 87}, 155149 (2013) [arXiv:1302.0847].
\bibitem{Festuccia&Liu}
G. Festuccia and H. Liu, 
Adv. Sci. Lett. {\bf 2}, 221 (2009) [arXiv:0811.1033].
\bibitem{Morgan&al}
J. Morgan, V. Cardoso, A. S. Miranda, C. Molina, and V. T. Zanchin, 
Phys. Rev. D {\bf 80}, 024024 (2009) [arXiv:0906.0064].
\bibitem{Fuini&al}
J. F. Fuini, C. F. Uhlemann, and L. G. Yaffe, 
JHEP {\bf 1612}, 042 (2016) [arXiv:1610.03491].
\bibitem{Horowitz&Hubeny}
G. T. Horowitz and V. E. Hubeny,
Phys. Rev. D {\bf 62}, 024027 (2000) [hep-th/9909056].
\bibitem{Nunez&Starinets} A. N\'{u}\~{n}ez and and A. O. Starinets,
Phys. Rev. D {\bf 67}, 124013 (2003) [hep-th/0302026].
\bibitem{BBH} S. Khlebnikov, 
Phys. Rev. D {\bf 75}, 065021 (2007) [hep-ph/0701043].
\bibitem{shock}
S. Khlebnikov, M. Kruczenski, and G. Michalogiorgakis, 
Phys. Rev. D {\bf 82}, 125003 (2010) [arXiv:1004.3803].
\bibitem{Fisher&al} M. P. A. Fisher, P. B. Weichman, G. Grinstein, and D. S. Fisher, 
  Phys. Rev. B {\bf 40}, 546 (1989).
\bibitem{MCStudy}
B. Capogrosso-Sansone, \c{S}. S\"{o}yler, N. Prokof'ev, and B. Svistunov,
Phys. Rev. A {\bf 77}, 015602 (2008) [arXiv:0710.2703].


\end{thebibliography}
\end{document}